\documentclass{article}%
\usepackage{amsmath}
\usepackage{amsfonts}
\usepackage{amssymb}
\usepackage{graphicx}%
\newcommand{\bpartial}{\mathop{\partial\kern -4pt\raisebox{.8pt}{$|$}}}
\newcommand{\bra}{\mathopen{[\kern-1.6pt[}}
\newcommand{\ket}{\mathclose{]\kern-1.5pt]}}
\newcommand{\bbra}{\mathopen{[\kern-2.2pt[\kern-2.3pt[}}
\newcommand{\bket}{\mathclose{]\kern-2.1pt]\kern-2.3pt]}}

\makeindex
\begin{document}
\title{\bf Algebraic Structures of N=(4,4) and  N=(8,8) SUSY Sigma Models on Lie groups and SUSY WZW Models  }
\author { M. Aali-Javanangrouh$^{a}$
\hspace{-2mm}{ \footnote{ e-mail:
aali@azaruniv.edu}}\hspace{1mm},\hspace{1mm} A.
Rezaei-Aghdam $^{a}$ \hspace{-2mm}{ \footnote{Corresponding author. e-mail:rezaei-a@azaruniv.edu }}\hspace{2mm}\\
{\small{\em
$^{a}$Department of Physics, Faculty of science, Azarbaijan Shahid Madani University }}\\
{\small{\em  53714-161, Tabriz, Iran  }}}

\maketitle
\begin{abstract}
Algebraic structures of  $N=(4,4)$ and $N=(8,8)$ supersymmetric (SUSY) two dimensional sigma models on Lie groups (in general) and SUSY Wess-Zumino-Witten (WZW) models (as special) are obtained. For SUSY WZW models, these algebraic structures are reduced to  Lie bialgebraic structures as for the $N=(2,2)$ SUSY WZW case; with the difference that there is a one 2-cocycle for the $N=(4,4)$ case and there are two  2-cocycles for the $N=(8,8)$ case. In general, we show that $N=(8,8)$ SUSY structure on Lie algebra  must be constructed from two $N=(4,4)$ SUSY structures and in special there must be  two 2-cocycles for Manin triples (one 2-cocycle for each of the $ N=(4,4) $ structures). Some examples are investigated. In this way, a calculational method for classifying the $N=(4,4)$ and $N=(8,8)$ structures on Lie algebras and Lie groups are obtained.
\end{abstract}
\section{\bf Introduction}
Supersymmetric two dimensional nonlinear sigma models have important role in theoretical and mathematical physics such as their numerous string applications. Let us have a short bibliography for this subject. The relation between these theories and geometry of the target spaces have been studied  about thirty five years ago \cite{bz}. The bi-Hermitean  geometry of the target spaces of the $N=2$ extended supersymmetric sigma models was first realized  in \cite{Gat} (see also \cite{hs}). Geometric condition of $N=(p,q)$ supersymmetric two dimensional nonlinear sigma models have been investigated in  \cite{ch} (see also \cite{PG}). Full discription of  hyper-k\"{a}hlerian  structure with torsion (HKT) (sigma models with $N=(4,0)$ ) and octonionic k\"{a}hler with torsion (OKT) have been studied  in \cite{Pap}. Sigma models with off-shell $N=(4,4)$ have been defined in \cite{Lin2}, in this case superfields of sigma models are chiral and twisted chiral. The extended supersymmetric sigma models on Lie group manifolds and also SUSY WZW models have been studied in \cite{sev}.  $N=2$ and $N=4$ extended superconformal field theories in two dimensions and also their correspondence with Manin triples have been investigated in \cite{Par1} and \cite{Par2} respectively (see also \cite{G}, \cite{Lin}). Also there are some notes about $N=8$ superconformal field theory in \cite{Par1}. The algebraic study of $N=(2,2)$ SUSY WZW models and also $N=(2,2)$ SUSY sigma models on Lie groups (\emph{algebraic bi-Hermitian structures}) have been studied in \cite{Lin} and \cite{RS} respectively. In \cite {bz}, the authors have argued that SUSY sigma model with $N>4$ is impossible, but in \cite{sev} it has been shown that it is possible on the Wolf spaces. In this work, we show that it is possible to have a SUSY sigma model on Lie group with $ N=8 $ structure, if  we have two independent  $ N=4 $ structure in the space and define it as a  double-bihypercomplex geometry. Here, we study  the algebraic structures of the $N=(4,4)$ and $N=(8,8)$ SUSY sigma models on Lie groups and also SUSY WZW models \footnote {Note that the algebraic structure for $p\neq q$  is algebaric structure with different Manin triples that we investigated in the later works.}. In other words, we will obtain a calculational method for classifying algebraic $N=(4,4)$ and $N=(8,8)$ structures on Lie algebras and Lie groups.

The outline of the paper is as follows: in section two, we review the $N=(2,2)$ SUSY sigma models on Lie groups and their \emph{algebraic bi-Hermitian structures}
 \cite{RS} as well as SUSY WZW models and their correspondence with Manin triples. Then, in section three, we obtain the \emph{algebraic bihypercomplex
 structures} for the $N=(4,4)$ SUSY sigma models on Lie groups and specially for the $N=(4,4)$ SUSY WZW models. We show their correspondence to Lie bialgebra with one 2-cocycle, furthermore at the end of this section we give an example for each case. Finally, in section four the algebraic structure of the $N=(8,8)$ two dimensional SUSY sigma
  models on Lie group is investigated. In general, we show that the $N=(8,8)$ structure on Lie group  is constructed from two $N=(4,4)$ structures
  and in the special case  for the $N=(8,8)$ SUSY WZW models algebraic structures are the Manin triples with two 2-cocycles, such that it is constructed from
  two Manin triples each with one 2-cocycle (two $N=(4,4)$ SUSY WZW structures) as stated but not proved in \cite{Par1}. An example is given at the end of this section.

\section{\bf  $N=(2,2)$ SUSY sigma models on Lie groups and SUSY WZW models }
In this section, for self contiaing of the paper we will review briefly the geometric description of the $N=(2,2)$ SUSY WZW and sigma models on Lie groups \cite{Gat}-\cite{sev} and their algebraic structures \cite{Lin},\cite{RS}. We will use the $N=(1,1)$ action  to the description of $N=(2,2)$ model; and impose extended supersymmetry on the superfields. With the knowledge that  $N$ supersymmetric sigma-models have $N$ supersymmetric generators $(Q_i)$ and $N-1$ complex structures $(J_i)$  on manifolds $M$ such that for $N=(p,q)$  SUSY sigma-models in two-dimension we  have $p$ right-handed generators $(Q_{i+})$   and $q$ left-handed generators$(Q_{i-})$ respectively,  $N=(1,1)$ SUSY sigma model has one right-handed generator $(Q_{+})$ and one left-handed generator $(Q_{-})$ and the action of the system on the manifold $M$ is written as follows \cite{Gat}:

\begin{equation}
S=\int d^{2}\sigma d^{2}\theta
D_{+}\Phi^{\mu}D_{-}\Phi^{\nu}(G_{\mu\nu}(\Phi)+B_{\mu\nu}(\Phi)),
\label{1}
\end{equation}
such that this action is invariant under the following supersymmetry transformation:
\begin{equation}
\delta_{1}(\epsilon)\Phi^{\mu}=i(\epsilon^{+}Q_{+}+\epsilon^{-}Q_{-})\Phi^{\mu},
\label{2}
\end{equation}
where $\Phi^{\mu}$ are $N=1$ superfields; so that their bosonic parts are the coordinates of the manifold $M$. Furthermore the bosonic parts of the $G_{\mu\nu}(\Phi)$ and $B_{\mu\nu}(\Phi)$ are metric and antisymmetric tensors on  $M$ respectively. Note that in the above relations $D_{\pm}$ are  superderivatives, and $\epsilon^{\pm}$ are parameters of supersymmetry transformations. The above action has also invariant under the following extended supersymmetry transformation \cite{Gat}:
\begin{equation}
\delta_{2}(\epsilon)\Phi^{\mu}=\epsilon^{+}D_{+}\Phi^{\nu}J^{\mu}_{+\nu}(\Phi)+\epsilon^{-}D_{-}\Phi^{\nu}J^{\mu}_{-\nu}(\Phi),
\label{3}
\end{equation}
where  $J_{\pm\nu}\hspace{0cm}^{\mu}\in TM\otimes T^{\ast}M$. The consequences of invariance of the action (\ref{1}) under the above transformations are the following conditions on $J_{\pm\sigma}\hspace{0cm}^{\rho}$\cite{Gat}:
\begin{eqnarray}
\label{complex}
J_{\pm\mu}\hspace{0cm}^{\lambda}J_{\pm\lambda}\hspace{0cm}^{\nu}&=&-\delta_{\mu}\hspace{0cm}^{\nu},
\\
\label{hermiti}
\hspace{1cm}J_{\pm  \rho}\hspace{0cm}^{\mu}G_{\mu\nu}&=&-G_{\mu\rho}J_{\pm  \nu}\hspace{0cm}^{\mu},
\\
\label{derivativ}
\nabla^{(\pm)}_{\rho}J_{\pm\nu}\hspace{0cm}^{\mu}&=&J_{\pm\nu,\rho}\hspace{0cm}^{\mu}+\Gamma^{\pm\mu}_{\rho\sigma}J_{\pm\nu}\hspace{0cm}^{\sigma}-\Gamma^{\pm\sigma}_{\rho\nu}J_{\pm\sigma}\hspace{0cm}^{\mu}=0,
\end{eqnarray}
where the extended connections $\Gamma^{\pm\mu}_{\rho\sigma}$ have the following forms:

\begin{equation}
\Gamma^{\pm\mu}_{\rho\nu}=\Gamma^{\mu}_{\rho\nu}\pm
G^{\mu\sigma}H_{\sigma\rho\nu},
\end{equation}
such that
\begin{equation}
H_{\mu\rho\sigma}=\frac{1}{2}(B_{\mu\rho,\sigma}+B_{\rho\sigma,\mu}+B_{\sigma\mu,\rho}),
\end{equation}
and $\Gamma^{\mu}_{\rho\nu}$ are Christofel symbols.
$$
$$
 In order to have a closed supersymmetry algebra we must have the integrability condition on the complex structures $(J_{\pm})$ (4) as follows \cite{Gat}:

\begin{equation}
N^{\rho}_{\mu\nu}(J_{\pm})=J_{\pm\mu}\hspace{0cm}^{\gamma}\partial_{[\gamma}J_{\pm\nu]}\hspace{0cm}^{\rho}-J_{\pm\nu}\hspace{0cm}^{\gamma}\partial_{[\gamma}J_{\pm\mu]}\hspace{0cm}^{\rho}=0.
\label{6}
\end{equation}

In this manner the $N=(2,2)$ SUSY structure of the sigma model on $M$ is equivalent to the existence of the  bi-Hermitian complex structure $(J_{\pm})$ on $M$ (\ref{complex}),(\ref{hermiti}),(\ref{6}) such that their covariant derivatives with respect to extended connection $\Gamma^{\pm\mu}_{\rho\nu}$ are equal to zero (\ref{derivativ}). If $M$ is a Lie group $G$ then in the non-coordinate bases, we have:
\begin{eqnarray}
\label{7}
G_{\mu\nu}&=&L_{\mu}\hspace{0cm}^{A}L_{\nu}\hspace{0cm}^{B}G_{AB}=R_{\mu}\hspace{0cm}^{A}R_{\nu}\hspace{0cm}^{B}G_{AB},
\\
\label{8}
f_{AB}\hspace{0cm}^{C}&=&L_{\nu}\hspace{0cm}^{C}(L^{\mu}\hspace{0cm}_{A}\partial_{\mu}L^{\nu}\hspace{0cm}_{B}-L^{\mu}\hspace{0cm}_{B}\partial_{\mu}L^{\nu}\hspace{0cm}_{A})=-
R_{\nu}\hspace{0cm}^{C}(R^{\mu}\hspace{0cm}_{A}\partial_{\mu}R^{\nu}\hspace{0cm}_{B}-R^{\mu}\hspace{0cm}_{B}\partial_{\mu}R^{\nu}\hspace{0cm}_{A}),
\\
\label{9}
J_{-\nu}\hspace{0cm}^{\mu}&=&L_{\nu}\hspace{0cm}^{B}J_{B}\hspace{0cm}^{A}L^{\mu}\hspace{0cm}_A,\hspace{2cm}J_{+\nu}\hspace{0cm}^{\mu}=R_{\nu}\hspace{0cm}^{B}J_{B}\hspace{0cm}^{A}R^{\mu}\hspace{0cm}_A,
\\
\label{10}
H_{\mu\nu\lambda}&=&L_{\mu}\hspace{0cm}^{A}L_{\nu}\hspace{0cm}^{B}L_{\lambda}\hspace{0cm}^{C}H_{ABC}=R_{\mu}\hspace{0cm}^{A}R_{\nu}\hspace{0cm}^{B}R_{\lambda}\hspace{0cm}^{C}H_{ABC},
\end{eqnarray}
where $G_{AB}$ is the ad-invariant nondegenerat metric and $H_{ABC}$ is antisymmetric tensor on the Lie algebra ${\bf g}$ of the Lie group $G$. Note that $L_{\mu}\hspace{0cm}^{A}(R_{\mu}\hspace{0cm}^{A})$ and $L^{\mu}\hspace{0cm}_A (R^{\mu}\hspace{0cm}_A)$ are components of left(right) invariant one-forms and their inverses on the Lie group $G$; $f_{AB}\hspace{0cm}^{C}$ are structure constants of the Lie algebra ${\bf g}$ and $J_B\hspace{0cm}^A$ is an algebraic map $J:{\bf g}\longrightarrow {\bf g}$ or \emph{algebraic complex structure}. Now, using the above relations and the following relations for the covariant derivative of the left invariant veilbin \cite{Na}:
\begin{equation}
\nabla^{\rho}L^{\eta}\hspace{0cm}_{A}=-\frac{1}{2}[f_{A}^{(\rho\eta)} + f^{\eta\rho}_{A}-T_{A}^{(\rho\eta)}-T^{\eta\rho}_{A}-L^{\eta B}\nabla^{\rho}G_{BA}+L^{\rho B}\nabla^{\eta}G_{AB}+L^{\eta}\hspace{0cm}_{A} L^{\rho}\hspace{0cm}_{B}\nabla_{A} G^{AB}],
\end{equation}
 we have the following algebraic relations, for the bi-Hermitian geometry of the $N=(2,2)$ SUSY sigma models on Lie group\cite{RS}:
\begin{eqnarray}
\label{metric}
G\chi_A&=&-(G\chi_A)^t,
\\
\label{J}
J_{C}\hspace{0cm}^{B}J_{B}\hspace{0cm}^{A}&=&-\delta_{C}\hspace{0cm}^{A},
\\
\label{JJ}
J_{C}\hspace{0cm}^{A}G_{AB}J^{B}\hspace{0cm}_{D}&=&G_{CD},
\\
\label{HH}
H_{EFG}&=&J_{E}\hspace{0cm}^{A}J_{F}\hspace{0cm}^{C}H_{ACG}+J_{G}\hspace{0cm}^{A}J_{E}\hspace{0cm}^{C}H_{ACF}+J_{F}\hspace{0cm}^{A}J_{G}\hspace{0cm}^{C}H_{ACE},
\\
\label{H}
(H_{A}+\chi_{A}G)J&=&[(\chi_{A}G+H_{A})J^{t}]^{t},
\end{eqnarray}
where $(\chi_{A})_{B}\hspace{0cm}^{C}=-f_{AB}\hspace{0cm}^{C}$ are the matrices in the adjoint representation and we have $(H_A)_{BC}=H_{ABC}$  for the matrices $H_A$. Note that relation (15) represents the ad-invariance of the Lie algebra metric $G_{AB}$. One can use relation (\ref{metric})-(\ref{H}) as a definition of \emph{algebraic bi-Hermitian structure} on Lie algebra \cite{RS}; and calculate and also classify such structures on the Lie algebras \cite{RS}. From (\ref{J}) we obtain the determinant of $J^{2}$ as $(-1)^{n}$, i.e. the dimension of the Lie algebra ${\bf g}$,  $n$ must be even and $J_{B}\hspace{0cm}^{A}$ has eigenvalues $\pm i$.
\\
\par
For the $N=(2,2)$ SUSY WZW models we have $H_{ABC} =f_{ABC} $ the relation  (\ref{H}) automatically is satisfied. If we choose a basis $T_A =(T_a ,T^{\bar{a}} )$ for the Lie algebra ${\bf g}$ then  we will have \cite{Lin}:
\begin{equation}
J=\left(\begin{array}{cc}
        i\delta^{a}_{b} & 0 \\
        0 & -i\delta^{\bar{a}}_{\bar{b}}
      \end{array}\right),
 \end{equation}
 where this form of $J$ and $H$  are  satisfied  in (\ref{HH}).  In this basis  according to (\ref{JJ}) we must have  the following form for $G_{AB}$:
  \begin{equation}
  G=\left(\begin{array}{cc}
        0 & g \\
         g^{'} & 0
      \end{array}\right),
       \end{equation}
 where $g$ and $g^{'}$ are a $\frac{n}{2}\times \frac{n}{2}$ symmetric matrix. According to (\ref{HH}), we have $f_{abc}=0$ and $f^{\bar{a}\bar{b}\bar{c}}=0$, this means that $f_{ab}^{\bar{c}}=f^{\bar{a}\bar{b}}_c=0$ i.e ${T_a }$ and ${T^{\bar{a}}}$ form Lie subalgebras ${\bf g_{+} }$ and ${\bf g_{-}}$ such that $({\bf g_+ }, {\bf g_-})$ is a Lie bialgebra and $({\bf g},{\bf g_+} ,{\bf g_-})$ is a Manin triple \cite{Lin}. The relation between Manin triples and $N=2$ superconformal models (from the algebraic OPE point of view) was first pointed out in \cite{Par1}. Also the relation of $N=(2,2)$ WZW models and Manin triple (from the action point of view) was shown in \cite{Lin}. In \cite{RS}, we have obtained all algebraic bi-Hermitian  structures related to four dimensional  real Lie algebra. Let us consider a simple example  for $N=(2,2)$ SUSY WZW models corresponding to the following non-Abelian four dimensional Manin triple ${\bf  A_{4,8}}$ \cite{RS}:
 \begin{equation}
 [T_2 ,T_4]=T_2,\hspace{1cm} [T_3 ,T_4]=-T_3,\hspace{1cm}[T_2 ,T_3]=T_1,
\end{equation}

\begin{equation}
G=\left(\begin{tabular}{cccc}
        0 & 0 & 0 & $1$ \\
        0 & 0 & $-1$ & 0 \\
        0 & $-1$ & 0 & 0 \\
        $1$ & 0 & 0 & 0 \\

      \end{tabular}\right),
\hspace{2cm}
J=\left(\begin{tabular}{cccc}
        0 & 1 & 0 & 0 \\
        -1 & 0 & 0 & 0 \\
        0 & 0 & 0 & 1 \\
        0 & 0 & -1 & 0 \\

      \end{tabular}\right).
\end{equation}
\section{\bf N=(4,4)  SUSY WZW and sigma models on Lie groups}
As mentioned above the correspondence between $N=2$,  $N=4$ and $N=8$ superconfomal Kac-Mody algebra and Manin triples have been investigated in \cite{Par1} (see also \cite{G}) and the Manin triples construction of $N=4$ superconformal field theories have also been investigated in \cite{Par2}, but up to now the algebraic structures of  the $N=(4,4)$ and $N=(8,8)$ SUSY sigma models on Lie groups and also   $N=(4,4)$ and $N=(8,8)$ SUSY WZW models and their relations to Manin triples (from the action point of view) are not studied explicitly. Here, in this section we will consider $ N=(4,4)$ case and in the next section will consider $ N=(8,8)$  case.

As in the previous section we consider the $ N=(1,1)$ SUSY sigma model action (\ref{1}) which is invariant under transformation (\ref{2}), but for $ N=(4,4)$ case one must impose the invariance of that action under the following extended SUSY transformations \cite{Gat},\cite{hs} (instead of (\ref{3})) \footnote { Note that for $ N=(4,4)$ SUSY sigma model we have four right-handed generators $(Q_{+r})$ and four left-handed generators $(Q_{-r})$ and three complex structures $(J_{\pm r})$.}.
\begin{equation}
\delta_{2r}(\epsilon)\Phi^{\mu}=\epsilon^{+}_{r}D_{+}\Phi^{\nu}J_{+r\nu}\hspace{0cm}^{\mu}(\Phi)+\epsilon^{-}_{r}D_{-}\Phi^{\nu}J_{-r\nu}\hspace{0cm}^{\mu}(\Phi),\hspace{2cm}r=1,2,3,
\end{equation}
such that the constraints on the complex structures are obtained as follows \cite{PG}:
\begin{eqnarray}
\label{18}
&&J_{\pm r\nu}\hspace{0cm}^{\lambda}J_{\pm r\lambda}\hspace{0cm}^{\mu}=-\delta^{\mu}_{\nu},
\\
\label{20}
&&J_{\pm r \rho}\hspace{0cm}^{\mu}G_{\mu\nu}=-G_{\mu\rho}J_{\pm r \nu}\hspace{0cm}^{\mu},\hspace{1cm}r=1,2,3,
\\
\label{21}
&&\nabla^{(\pm)}_{\rho}J_{\pm r \nu}\hspace{0cm}^{\mu}=\partial_{\rho}J_{\pm r \nu}\hspace{0cm}^{\mu}+\Gamma^{\pm\mu}_{\rho\sigma}J_{\pm r \nu}\hspace{0cm}^{\sigma}-\Gamma^{\pm\sigma}_{\rho\nu}J_{\pm r \sigma}\hspace{0cm}^{\mu}=0,\hspace{1cm}r=1,2,3.
\end{eqnarray}
From the fact that the algebra of SUSY transformations must be closed (i.e $[\delta_{r}^{2}(\epsilon_{r}),\delta_{r}^{2}(\epsilon_{r})]$,
 \\
and $[\delta_{r}^{2}(\epsilon_{r}),\delta_{s}^{2}(\epsilon_{s})]$) the following relations are obtained  \cite{PG}:
\begin{eqnarray}
\label{22}
&&J_{\pm r}\hspace{0cm}^{\lambda}\hspace{0cm}_{[\mu}\partial_{\lambda}J_{\pm r}\hspace{0cm}^{\gamma}\hspace{0cm}_{\nu]}-J_{\pm r}\hspace{0cm}^{\lambda}\hspace{0cm}_{[\mu}\partial_{\lambda}J_{\pm r}\hspace{0cm}^{\gamma}\hspace{0cm}_{\nu]}=0,\hspace{1cm}r=1,2,3,
\\
\label{23}
&&J_{\pm r}\hspace{0cm}^{\mu}\hspace{0cm}_{\lambda}J_{\pm s}\hspace{0cm}^{\lambda}\hspace{0cm}_{\nu}+J_{\pm s}\hspace{0cm}^{\mu}\hspace{0cm}_{\lambda}J_{\pm r}\hspace{0cm}^{\lambda}\hspace{0cm}_{\nu}=0,\hspace{3cm} r\neq s,
\\
\label{forget}
&&J_{\mp r}\hspace{0cm}^{\mu}\hspace{0cm}_{\lambda}J_{\pm s}\hspace{0cm}^{\lambda}\hspace{0cm}_{\nu}+J_{\mp s}\hspace{0cm}^{\mu}\hspace{0cm}_{\lambda}J_{\pm r}\hspace{0cm}^{\lambda}\hspace{0cm}_{\nu}=0,\hspace{3cm} r\neq s,
\\
\label{24}
&&J_{\pm r}\hspace{0cm}^{\gamma}\hspace{0cm}_{\lambda}\partial_{[v}J_{\pm s}\hspace{0cm}^{\lambda}\hspace{0cm}_{\mu]}+J_{\pm r}\hspace{0cm}^{\lambda}\hspace{0cm}_{[\mu}\partial_{\lambda}J_{\pm s}\hspace{0cm}^{\gamma}\hspace{0cm}_{\nu]}+J_{\pm s}\hspace{0cm}^{\gamma}\hspace{0cm}_{\lambda}\partial_{[v}J_{\pm r}\hspace{0cm}^{\lambda}\hspace{0cm}_{\mu]}+J_{\pm s}\hspace{0cm}^{\lambda}\hspace{0cm}_{[\mu}\partial_{\lambda}J_{\pm r}\hspace{0cm}^{\gamma}\hspace{0cm}_{\nu]}=0,
\end{eqnarray}
such that these  are  Nijenhuis-concomitant\hspace{-1mm}{ \footnote{The Nijenhuis concomitant of $J_r$ and $J_s$ has the following form \cite{FN}:

 \begin{eqnarray}
 \nonumber
 N(I,J)^{\lambda}\hspace{0cm}_{\mu\nu}=[I^{\gamma}\hspace{0cm}_{\mu}\partial_{\gamma}J^{\lambda}\hspace{0cm}_{\nu}-(\mu\longleftrightarrow \nu)-(I^{\lambda}\hspace{0cm}_{\gamma}\partial_{\mu}J^{\gamma}\hspace{0cm}_{\nu}-(\mu\longleftrightarrow \nu))]+(I\longleftrightarrow J)
 \end{eqnarray}} \cite{FN} for complex structures $J_{\pm r}$.
 When the background is a Lie group $G$ then in  non-coordinate bases ((\ref{7})-(\ref{10})) the geometrical relations  (\ref{18})-(\ref{24}) have the following algebraic forms:

\begin{eqnarray}
\label{25}
J_{rC}\hspace{0cm}^{B}J_{rB}\hspace{0cm}^{A}&=&-\delta_{C}\hspace{0cm}^{A},
\\
\label{27}
J_{rC}\hspace{0cm}^{A}G_{AB}J_{rD}\hspace{0cm}^B&=&G_{DC},\hspace{1cm}r=1,2,3,
\\
\label{29}
(H_{A}+\chi_{A}G)J_r&=&[(\chi_{A}G+H_{A})J_r^{t}]^{t},\hspace{0.25cm}r=1,2,3,
\\
\label{30}
 J_{s D}\hspace{0cm}^{B}J_{r B}\hspace{0cm}^{A}+J_{r D}\hspace{0cm}^{B}J_{s B}\hspace{0cm}^{A}&=&0,\hspace{1cm} r\neq s,
\end{eqnarray}\vspace{-5mm}
\begin{eqnarray}
\label{28}
H_{EFG}=J_{rE}\hspace{0cm}^{A}J_{rF}\hspace{0cm}^{C}H_{ACG}+J_{rG}\hspace{0cm}^{A}J_{rE}\hspace{0cm}^{C}H_{ACF}&+&J_{rF}\hspace{0cm}^{A}J_{rG}\hspace{0cm}^{C}H_{ACE},\hspace{0.25cm}r=1,2,3,
\\
\nonumber
H_{B^{'}A}\hspace{0cm}^{B}J_{r B}\hspace{0cm}^{C^{'}}J_{s A^{'}}\hspace{0cm}^{A}+H_{A^{'}B^{'}}\hspace{0cm}^{A}J_{r B}\hspace{0cm}^{C^{'}}J_{s A}\hspace{0cm}^{B}+H_{AA^{'}}\hspace{0cm}^{B}J_{r B}\hspace{0cm}^{C^{'}}J_{sB^{'}}\hspace{0cm}^{A}&+&H_{AB}\hspace{0cm}^{C^{'}}J_{rA^{'}}\hspace{0cm}^{A}J_{sB^{'}}\hspace{0cm}^{B}+H_{B^{'}A}\hspace{0cm}^{B}J_{r A^{'}}\hspace{0cm}^{A}J_{sB}\hspace{0cm}^{C^{'}}
\\
\label{31}
+H_{BA}\hspace{0cm}^{C^{'}}J_{rB^{'}}\hspace{0cm}^{A}J_{sA^{'}}\hspace{0cm}^{B}+H_{AA^{'}}\hspace{0cm}^{B}J_{r B^{'}}\hspace{0cm}^{A}J_{s B}\hspace{0cm}^{C^{'}}&+&H_{A^{'}B^{'}}\hspace{0cm}^{B}J_{r B}\hspace{0cm}^{A}J_{s A}\hspace{0cm}^{C^{'}}=0.
\end{eqnarray}
In this way, the relations (\ref{25})-(\ref{31}) define the \emph{algebraic bihypercomplex structures}{ \footnote{Similar to the name of bihypercomplex geometry \cite{Lin1}. } on the Lie algebra ${\bf g}$, such that we have three algebraic complex structures $J_r ,  (r=1,2,3)$ where by use of (\ref{25}) and (\ref{30}) only two of them are independent i.e we have two algebraic independent  complex structures (e.g $J_1$ and $J_2$)\footnote{If we consider $ J_1.J_2=a_1 J_1 + a_2 J_2 +a_3 J_3 $ then by use of (\ref{25}) we conclude $ J_1.J_2=\pm J_3 $.}, also according to the relations (\ref{25}) the dimension of Lie algebra ${\bf g}$ must be $(4n)$ .
\par
Similar to  $N=(2,2)$ case for $N=(4,4)$ SUSY WZW models we have $H_{ABC} = f_{ABC}$. Then, the relations (\ref{29}) automatically are satisfied and from (\ref{25}),(\ref{27}) and (\ref{28}) one can obtain the following forms for $J_1$, $J_2$ and $G$:
\begin{eqnarray}
G=\left(\begin{array}{cc}
        0 & g \\
         g^{'} & 0
      \end{array}\right),
\hspace{2cm}J_{1}=\left(\begin{array}{cc}
        i\delta^{a}_{b} & 0 \\
        0 & -i\delta^{\bar{a}}_{\bar{b}}
      \end{array}\right),\hspace{2cm}
J_{2}=\left(\begin{array}{cc}
        R_{a}^{b} & R_{a\bar{b}} \\
        R^{\bar{a}b} & R^{\bar{a}}_{\bar{b}}
      \end{array}\right),
 \end{eqnarray}
  where  we use the basis $T_A = \{ T_a , T^{\bar{a}} \}$ for the Lie algebra ${\bf g}$. Then, from (\ref{30}) one can obtain $R_{a}^{b}=R^{\bar{a}}_{\bar{b}}=0$, and from (\ref{25}) $(R_{a\bar{b}})R^{\bar{a}b}=-1$. In this way, the $J_2$ has the following form:
 \begin{eqnarray}
 \label{form J2}
J_{2}=\left(\begin{array}{cc}
       0 & R_{a\bar{b}} \\
       -(R_{a\bar{b}})^{-1} & 0
      \end{array}\right).
 \end{eqnarray}
  Note that from (\ref{28}) as for $N=(2,2)$ case we see that ${\bf g}={\bf g_+}\oplus{\bf g_-}$ where ${\bf g_+}$ and ${\bf g_-} $ are Lie subalgebras with basis $T_\Gamma =\{T_a,T^{\bar{a}}\}$ and $T_\Gamma ^{'}=\{T_{a}^{'},T^{'\bar{a}}\}$, $\bar{a},a=1,....,n$, such that the basis for ${\bf g}$ are now $T_A = \{T_{\Gamma} , T_{\bar{\Gamma}}\}$ ,i.e. they form a Lie bialgebra, so that from (\ref{31}) we have:
\begin{equation}
 f_{AB}\hspace{0cm}^{D} R_{DC}+f_{BC}\hspace{0cm}^{D}R_{DA}-f_{CA}\hspace{0cm}^{D} R_{DB}=0.
\label{32}
\end{equation}
This means that we have a 2-cocycle\footnote{To show this, we consider the definition of coboundary operator $\delta$ on an \emph{i}-cochain $\gamma$ on the Lie algebra $
{\bf g}$ with values in the space $M$ as follows \cite{SS}:
\begin{equation}
\nonumber
\delta \gamma(T_{0},T_{1},...,T_{i})=\Sigma_{j=0}^{i} T_{j}\otimes( \gamma(T_{0},...,\hat{T}_{j},...,T_{i}))+\Sigma_{j<k}(-1)^{j+k} \gamma([T_{j},T_{k}],T_{0},...,\hat{T}_{j},...,T_{k},...,T_{i}),
\label{33}
\end{equation}
$\forall T_A \in {\bf g}$. The two cochain $\gamma$ is two cocycle when $\delta\gamma =0$. Now, for the case  $M= \mathbb{C}$ we have:
\begin{eqnarray}
\nonumber
-\delta \gamma(T_{0},T_{1},T_{2})&+& T_{0}\otimes( \gamma(T_{1},T_{2}))+T_{1}\otimes( \gamma(T_{0},T_{2}))+T_{2}\otimes( \gamma(T_{0},T_{1}))
\\
\nonumber
&-& \gamma([T_{0},T_{1}],T_{2})+ \gamma([T_{0},T_{2}],T_{1})- \gamma([T_{1},T_{2}],T_{0})=0.
\label{34}
\end{eqnarray}
 Using the following form for the two cochain:
 \begin{equation}
\nonumber
\gamma (T_A,T_B)=(R_{AB})^{\Gamma\Lambda} T_\Gamma \otimes T_\Lambda +(R_{AB})^{\Gamma \bar{\Lambda}} T_\Gamma \otimes T_{\bar{\Lambda}} +(R_{AB})^{\bar{\Gamma}\Lambda} T_{\bar{\Gamma}} \otimes T_\Lambda +(R_{AB})^{\bar{\Gamma}\bar{\Lambda}} T_{\bar{\Gamma}} \otimes T_{\bar{\Lambda}},
\label{35}
\end{equation}
 after some calculation one can obtain (\ref{32}).}.  In this way the algebraic structure of $N=(4,4)$ SUSY WZW models is also Lie bialgebra as for the $N=(2,2)$ SUSY WZW models with the difference that for the $N=(4,4)$ case, we have Lie bialgebra with a 2-cocycle, such that independent algebraic complex structures $(J_1,J_2)$ are anticommuting (\ref{30}).
\subsection{Examples}
a) As an  example for  $N=(4,4)$  SUSY sigma models on Lie group  we consider the  four dimensional  Lie group ${\bf IX+R}$  \cite{P}.  For this example, one can find from (\ref{25})-(\ref{31}) the following forms for the metric $G$ and complex structures $J_1$ and $J_2$:
\begin{eqnarray}
G=\left(\begin{array}{cccc}
  \beta & 0 & 0 & 0 \\
  0 & \beta & 0 & 0 \\
  0 & 0 & \beta & 0 \\
  0 & 0 & 0 & \beta
\end{array}\right),
\hspace{1cm}
J_1=\left(\begin{array}{cccc}
  0 & -1& 0 & 0 \\
  1 & 0 & 0 & 0 \\
  0 & 0 & 0 & 1 \\
  0 & 0 & -1 & 0
\end{array}\right),\hspace{1cm}J_2=\left(\begin{array}{cccc}
  0 & 0 & 0 & -1 \\
  0 & 0 & 1 & 0 \\
  0 & -1 & 0 & 0 \\
  1 & 0 & 0 & 0
\end{array}\right),
\end{eqnarray}
and for $H$s we have:
\begin{eqnarray}
H_1=\left(\begin{array}{cccc}
 0 & 0 & 0 & 0 \\
  0 & 0 & \beta & 0 \\
  0 & -\beta & 0 & 0 \\
  0 & 0 & 0 & 0
\end{array}\right),
\hspace{0cm}
H_2=\left(\begin{array}{cccc}
  0 & 0& -\beta & 0 \\
  0 & 0 & 0 & 0 \\
  \beta & 0 & 0 & 0 \\
  0 & 0 & 0 & 0
\end{array}\right),\hspace{0cm}H_3=\left(\begin{array}{cccc}
  0 & 0 & 0 & 0 \\
  0 & 0 & 0 & 0 \\
  0 & 0 & 0 & 0 \\
  0 & 0 & 0 & 0
\end{array}\right),\hspace{0cm}H_4=\left(\begin{array}{cccc}
  0 & \beta & 0 & 0 \\
  -\beta & 0 & 0 & 0 \\
  0 & 0 & 0 & 0 \\
  0 & 0 & 0 & 0
\end{array}\right).
\end{eqnarray}
\par
b) As an example for the $N=(4,4)$ SUSY WZW model  we have obtained the non-Abelian eight dimensional bi-Hermitian structure on Poisson-Lie group  $\bf D$ related to the bialgebra ($ A_{4,5}^{-1,-1}\oplus II+R$) \footnote{The algebras $ A_{4,5}^{-1,-1}$ and $ II+R$ are four dimensional Lie algebras \cite{P} such that they are dual to each other (i.e. four a Lie bialgebra \cite{D}).} with the following commutation relations for their Lie bialgebra:
\begin{eqnarray}
 \nonumber
[T_{1},T_{4}]&=& T_{1},\hspace{1cm}[T_{2},T_{4}]=-T_{2},\hspace{1cm}[T_{3},T_{4}]=-T_{3},\hspace{1cm}[T_{4},T_{5}]=T_{5}\\
\nonumber
[T_{2},T_{6}]&=& T_{7},\hspace{1cm}[T_{1},T_{5}]=-T_{8},\hspace{1cm}[T_{3},T_{6}]=-T_{6},\hspace{1cm}[T_{3},T_{7}]=T_{8},\hspace{1cm}[T_{4},T_{7}]=-T_{7}.
\end{eqnarray}
 where the $ \{T_{1},...,T_{4}\} $  and $ \{T_{5},...,T_{8}\} $ are the base of the Lie algebra $ { A_{4,5}^{-1,-1}} $ and $ { II+R} $ respectively.
From (\ref{25})-(\ref{31}) one can obtain the following forms for the metric $G$ and complex structures $J_1$ and $J_2$.
\begin{eqnarray}
G=\left (\begin {array}{cc} 0&I_{4*4}\\
I_{4*4}&0\\ \end {array} \right),J_1=\left (\begin {array}{cc} A&0\\
0&A\\ \end {array} \right),J_2=\left (\begin {array}{cc} 0&R\\
-R^{-1}&0\\ \end {array} \right),
\end{eqnarray}
where
\begin{eqnarray}
A=\left(\begin{array}{cccc}
  0 & 0 & 0 & 1 \\
  0 & 0 & -1 & 0\\
  0 & 1 & 0 & 0\\
  -1 & 0 & 0 & 0\\
\end{array}\right),\hspace{1cm}R=\left(\begin{array}{cccc}
  0 & -1 & 0 & 0 \\
  1 & 0 & 0 & 0\\
  0 & 0 & 0 & 1\\
  0 & 0 & -1 & 0\\
\end{array}\right).
\end{eqnarray}
\section{\bf N=(8,8) SUSY WZW and sigma models on Lie groups}
Now, as for the $N=(4,4)$ case we consider the action (\ref{1}) again;  such that this action is  invariant under SUSY transformation (\ref{2}) as well as  the following second SUSY transformations (instead of (\ref{3}))\cite{PG}:

\begin{equation}
\delta_{2r}(\epsilon)\Phi^{\mu}=\epsilon^{+}_{r}D_{+}\Phi^{\nu}J_{+r\nu}\hspace{0cm}^{\mu}(\Phi)+\epsilon^{-}_{r}D_{-}\Phi^{\nu}J_{-r\nu}\hspace{0cm}^{\mu}(\Phi),\hspace{2cm}r=1,...,7,
\label{trans}
\end{equation}
where for these transformations we have fourteen $J_{\pm r}$ geometric complex structures. As for  $N=(4,4)$ case from the invariance of the action (\ref{1}) under transformation (\ref{trans}) and the property that  the algebra of transformations must be closed, one can obtain again relations similar to (\ref{18})-(\ref{24}) with $(r=1,...,7)$ \cite{PG} and also the same algebraic relations (\ref{25})-(\ref{31}). For this case from (\ref{30}) and (\ref{25}) we have obtained the following relations between complex structures:
\begin{eqnarray}
\label{J1}
J_{1C}^BJ_{2B}^A&=&J_{3C}^A,\hspace{.5cm}J_{1C}^BJ_{4B}^A=J_{6C}^A,\hspace{.5cm}J_{1C}^BJ_{5B}^A=J_{7C}^A,
\\
\label{J2}
J_{2C}^BJ_{4B}^A&=&J_{7C}^A,\hspace{.5cm}J_{2C}^BJ_{5B}^A=J_{6C}^A,\hspace{.5cm}
J_{3C}^BJ_{4B}^A=J_{5C}^A,\hspace{.5cm}J_{3C}^BJ_{6B}^A=J_{7C}^A,
\end{eqnarray}
 such that from (\ref{J1}) four of the complex structures are independent i.e. $\{J_1,J_2,J_4,J_5\}$ where the  algebraic relations (\ref{25})-(\ref{31}) show that the two pairs $\{J_1,J_2\}$ and $\{J_4,J_5\}$ costruct two  $N=(4,4)$ SUSY structures, on the other hand  from (\ref{J2})  one of these four complex structures (e.g. $ J_5 $) are dependent.
\\
\par
 As for the $N=(4,4)$ case, for $N=(8,8)$ SUSY WZW we have obtained the following forms for the complex structures $J_1$, $J_2$ and $J_4$ and also for $G$:
  \begin{eqnarray}
G=\left(\begin{array}{cc}
        0 & g \\
         g^{'} & 0
      \end{array}\right)
      , \hspace{.25cm}J_{1}=\left(\begin{array}{cc}
        i\delta^{a}_{b} & 0 \\
        0 & -i\delta^{\bar{a}}_{\bar{b}}
      \end{array}\right)
, \hspace{.25cm}
  J_{2}=\left(\begin{array}{cc}
        0 & R_{1a\bar{b}} \\
       -( R_{1a\bar{b}})^{-1} & 0
      \end{array}\right),\hspace{.25cm} J_{4}=\left(\begin{array}{cc}
        0 & R_{2a\bar{b}} \\
       -( R_{2a\bar{b}})^{-1} & 0
      \end{array}\right).
 \end{eqnarray}
In this case relation (\ref{31}) reduces to the following relations:
 \begin{equation}
f_{AB}\hspace{0cm}^{D} R_{1D}\hspace{0cm}^{C}+f_{DA}\hspace{0cm}^{C} R_{1B}\hspace{0cm}^{D}-f_{BD}\hspace{0cm}^{C} R_{1A}\hspace{0cm}^{D}=0,
\end{equation}
\begin{equation}
f_{AB}\hspace{0cm}^{D} R_{2D}\hspace{0cm}^{C}+f_{DA}\hspace{0cm}^{C} R_{2B}\hspace{0cm}^{D}-f_{BD}\hspace{0cm}^{C} R_{2A}\hspace{0cm}^{D}=0,
\end{equation}
i.e the algebraic structures of $N=(8,8)$ SUSY WZW models are Lie bialgebras with two 2-cocycles. Such that  the Lie bialgebra is constructed from two Lie bialgebra related to two $N=(4,4)$ structure with $ \{J_1,J_2\} $ and  $ \{J_4,J_5\} $ each with one 2-cocycles so that  from four complex structures $ \{J_1,J_2 , J_4,J_5\} $ three of them  e.g. $J_1$, $J_2$ and $J_4$ are independent and  anticommute (\ref{30}). In this way, relations (\ref{25})-(\ref{31})  define the \emph{algebraic double-bihypercomplex structures} on the Lie group $\bf g$.
\subsection{Example}
As  an example for  $N=(8,8)$  SUSY sigma models on Lie group we consider the  eight dimensional Lie group  ${ IX+R \oplus IX+R }$ \cite{P}. Now, in this case  from  (\ref{25})-(\ref{31}) one can obtain the following forms for the metric $G$ and complex structures $J_1$ , $J_2$ and $J_3$:
\begin{eqnarray}
G=\left (\begin {array}{cc} I_{4*4}&0\\
0&I_{4*4}\\ \end {array} \right),J_1=\left (\begin {array}{cc} A&0\\
0&-A\\ \end {array} \right),J_2=\left (\begin {array}{cc} 0&R_1\\
R_1&0\\ \end {array} \right),J_3=\left (\begin {array}{cc} 0&R_2\\
-R_2&0\\ \end {array} \right),
\end{eqnarray}
where
\begin{eqnarray}
A=\left(\begin{array}{cccc}
  0 & 1 & 0 & 0 \\
  -1 & 0 & 0 & 0\\
  0 & 0 & 0 & 1\\
  0 & 0 & -1 & 0\\
\end{array}\right),\hspace{1cm}R_1=\left(\begin{array}{cccc}
  0 & 1 & 0 & 0 \\
  -1 & 0 & 0 & 0\\
  0 & 0 & 1 & 0\\
  0 & 0 & 0 & -1\\
\end{array}\right),\hspace{1cm}R_2=\left(\begin{array}{cccc}
  1 & 0 & 0 & 0 \\
  0 & 1 & 0 & 0\\
  0 & 0 & 0 & -1\\
  0 & 0 & -1 & 0\\
\end{array}\right),
\end{eqnarray}
and for  $H$s we have \footnote{The $ a_1,a_2,b_1,b_2,c_1,c_2,d_1,e_1,e_2,e_3,e_4,f_1,f_2,g_1,s_1,s_2,s_3,s_4,r_1,k_1,k_2,l_1,m_1,m_2,n_1,p_1,p_2,p_3,p_4,t_1,q_1$and $q_2$ are arbitrary real parameters.}:
\begin{eqnarray}
\nonumber
&&\hspace{-1cm}\scriptsize H_1=\left( \begin {array}{cc|cc|cc|cc} 0&0&0&0&0&0&0&0\\ \noalign{\medskip}0
&0&{\it a_1}&{\it a_2}&{\it b_1}&{\it b_2}&{\it c_1}&{\it c_2}
\\ \hline \noalign{\medskip}0&-{\it a_1}&0&{\it d_1}&{\it e_1}&{\it e_2}&{\it f_2}
&-{\it f_1}\\ \noalign{\medskip}0&-{\it a_2}&-{\it d_1}&0&{\it e_3}&{\it 
e_4}&{\it f_1}&{\it f_2}\\\hline \noalign{\medskip}0&-{\it b_1}&-{\it e_1}&-{\it 
e_3}&0&{\it g_1}&{\it s_1}&{\it s_2}\\ \noalign{\medskip}0&-{\it b_2}&-{
\it e_2}&-{\it e_4}&-{\it g_1}&0&{\it s_3}&{\it s_4}\\\hline \noalign{\medskip}0&
-{\it c_1}&-{\it f_2}&-{\it f_1}&-{\it s_1}&-{\it s_3}&0&-{\it d_1}
\\ \noalign{\medskip}0&-{\it c_2}&{\it f_1}&-{\it f_2}&-{\it s_2}&-{\it s_4
}&{\it d_1}&0\end {array} \right) 
  ,\scriptsize  H_2=\left( \begin {array}{cc|cc|cc|cc} 0&0&-{\it a_1}&-{\it a_2}&-{\it b_1}&-{
\it b_2}&-{\it c_1}&-{\it c_2}\\ \noalign{\medskip}0&0&0&0&0&0&0&0
\\\hline \noalign{\medskip}{\it a_1}&0&0&{\it r_1}&{\it e_2}&-{\it e_1}&{\it k_1}
&{\it k_2}\\ \noalign{\medskip}{\it a_2}&0&-{\it r_1}&0&{\it e_4}&-{\it e_3
}&-{\it k_2}&{\it k_1}\\\hline \noalign{\medskip}{\it b_1}&0&-{\it e_2}&-{\it e_4
}&0&{\it l_1}&{\it s_3}&{\it s_4}\\ \noalign{\medskip}{\it b_2}&0&{\it e_1}
&{\it e_3}&-{\it l_1}&0&-{\it s_1}&-{\it s_2}\\\hline \noalign{\medskip}{\it c_1}
&0&-{\it k_1}&{\it k_2}&-{\it s_3}&{\it s_1}&0&-{\it r_1}
\\ \noalign{\medskip}{\it c_2}&0&-{\it k_2}&-{\it k_1}&-{\it s_4}&{\it s_2}
&{\it r_1}&0\end {array} \right),
\\
\nonumber
&&\hspace{-1cm}\scriptsize H_3=\left( \begin {array}{cc|cc|cc|cc} 0&{\it a_1}&0&-{\it d_1}&-{\it e_1}&-{
\it e_2}&-{\it f_2}&{\it f_1}\\ \noalign{\medskip}-{\it a_1}&0&0&-{\it r_1}
&-{\it e_2}&{\it e_1}&-{\it k_1}&-{\it k_2}\\\hline \noalign{\medskip}0&0&0&0&0&0
&0&0\\ \noalign{\medskip}{\it d_1}&{\it r_1}&0&0&-{\it q_1}&-{\it q_2}&{
\it m_1}&{\it m_2}\\\hline \noalign{\medskip}{\it e_1}&{\it e_2}&0&{\it q_1}&0&{
\it a_1}&{\it p_2}&-{\it p_1}\\ \noalign{\medskip}{\it e_2}&-{\it e_1}&0&{
\it q_2}&-{\it a_1}&0&{\it p_4}&-{\it p_3}\\\hline \noalign{\medskip}{\it f_2}&{
\it k_1}&0&-{\it m_1}&-{\it p_2}&-{\it p_4}&0&{\it n_1}
\\ \noalign{\medskip}-{\it f_1}&{\it k_2}&0&-{\it m_2}&{\it p_1}&{\it p_3}&
-{\it n_1}&0\end {array} \right)
,\scriptsize H_4=\left( \begin {array}{cc|cc|cc|cc} 0&{\it a_2}&{\it d_1}&0&-{\it e_3}&-{
\it e_4}&-{\it f_1}&-{\it f_2}\\ \noalign{\medskip}-{\it a_2}&0&{\it r_1}&0
&-{\it e_4}&{\it e_3}&{\it k_2}&-{\it k_1}\\\hline \noalign{\medskip}-{\it d_1}&-
{\it r_1}&0&0&{\it q_1}&{\it q_2}&-{\it m_1}&-{\it m_2}
\\ \noalign{\medskip}0&0&0&0&0&0&0&0\\\hline \noalign{\medskip}{\it e_3}&{
\it e_4}&-{\it q_1}&0&0&{\it a_2}&{\it p_1}&{\it p_2}\\ \noalign{\medskip}{
\it e_4}&-{\it e_3}&-{\it q_2}&0&-{\it a_2}&0&{\it p_3}&{\it p_4}
\\ \hline\noalign{\medskip}{\it f_1}&-{\it k_2}&{\it m_1}&0&-{\it p_1}&-{\it p_3}
&0&{\it t_1}\\ \noalign{\medskip}{\it f_2}&{\it k_1}&{\it m_2}&0&-{\it p_2}
&-{\it p_4}&-{\it t_1}&0\end {array} \right),
\\
\nonumber
&&\hspace{-1cm}\scriptsize H_5=\left( \begin {array}{cc|cc|cc|cc} 0&{\it b_1}&{\it e_1}&{\it e_3}&0&-{\it 
g_1}&-{\it s_1}&-{\it s_2}\\ \noalign{\medskip}-{\it b_1}&0&{\it e_2}&{\it 
e_4}&0&-{\it l_1}&-{\it s_3}&-{\it s_4}\\\hline \noalign{\medskip}-{\it e_1}&-{
\it e_2}&0&-{\it q_1}&0&-{\it a_1}&-{\it p_2}&{\it p_1}
\\ \noalign{\medskip}-{\it e_3}&-{\it e_4}&{\it q_1}&0&0&-{\it a_2}&-{\it 
p_1}&-{\it p_2}\\ \hline\noalign{\medskip}0&0&0&0&0&0&0&0\\ \noalign{\medskip}
{\it g_1}&{\it l_1}&{\it a_1}&{\it a_2}&0&0&{\it c_1}&{\it c_2}
\\ \hline\noalign{\medskip}{\it s_1}&{\it s_3}&{\it p_2}&{\it p_1}&0&-{\it c_1}&0
&{\it q_1}\\ \noalign{\medskip}{\it s_2}&{\it s_4}&-{\it p_1}&{\it p_2}&0&-
{\it c_2}&-{\it q_1}&0\end {array} \right)
 ,\scriptsize H_6=\left( \begin {array}{cc|cc|cc|cc} 0&{\it b_2}&{\it e_2}&{\it e_4}&{\it g_1}
&0&-{\it s_3}&-{\it s_4}\\ \noalign{\medskip}-{\it b_2}&0&-{\it e_1}&-{
\it e_3}&{\it l_1}&0&{\it s_1}&{\it s_2}\\\hline \noalign{\medskip}-{\it e_2}&{
\it e_1}&0&-{\it q_2}&{\it a_1}&0&-{\it p_4}&{\it p_3}\\ \noalign{\medskip}
-{\it e_4}&{\it e_3}&{\it q_2}&0&{\it a_2}&0&-{\it p_3}&-{\it p_4}
\\\hline \noalign{\medskip}-{\it g_1}&-{\it l_1}&-{\it a_1}&-{\it a_2}&0&0&-{
\it c_1}&-{\it c_2}\\ \noalign{\medskip}0&0&0&0&0&0&0&0
\\\hline \noalign{\medskip}{\it s_3}&-{\it s_1}&{\it p_4}&{\it p_3}&{\it c_1}&0&0
&{\it q_2}\\ \noalign{\medskip}{\it s_4}&-{\it s_2}&-{\it p_3}&{\it p_4}&{
\it c_2}&0&-{\it q_2}&0\end {array} \right),
\\
\nonumber
&& \hspace{-1cm}\scriptsize H_7= \left( \begin {array}{cc|cc|cc|cc} 0&{\it c_1}&{\it f_2}&{\it f_1}&{\it s_1}
&{\it s_3}&0&{\it d_1}\\ \noalign{\medskip}-{\it c_1}&0&{\it k_1}&-{\it k_2
}&{\it s_3}&-{\it s_1}&0&{\it r_1}\\\hline \noalign{\medskip}-{\it f_2}&-{\it k_1
}&0&{\it m_1}&{\it p_2}&{\it p_4}&0&-{\it n_1}\\ \noalign{\medskip}-{\it 
f_1}&{\it k_2}&-{\it m_1}&0&{\it p_1}&{\it p_3}&0&-{\it t_1}
\\ \hline\noalign{\medskip}-{\it s_1}&-{\it s_3}&-{\it p_2}&-{\it p_1}&0&{\it c_1
}&0&-{\it q_1}\\ \noalign{\medskip}-{\it s_3}&{\it s_1}&-{\it p_4}&-{\it 
p_3}&-{\it c_1}&0&0&-{\it q_2}\\\hline \noalign{\medskip}0&0&0&0&0&0&0&0
\\ \noalign{\medskip}-{\it d_1}&-{\it r_1}&{\it n_1}&{\it t_1}&{\it q_1}&{
\it q_2}&0&0\end {array} \right)
 ,\scriptsize H_8=\left( \begin {array}{cc|cc|cc|cc} 0&{\it c_2}&-{\it f_1}&{\it f_2}&{\it s_2
}&{\it s_4}&-{\it d_1}&0\\ \noalign{\medskip}-{\it c_2}&0&{\it k_2}&{\it 
k_1}&{\it s_4}&-{\it s_2}&-{\it r_1}&0\\ \hline\noalign{\medskip}{\it f_1}&-{\it 
k_2}&0&{\it m_2}&-{\it p_1}&-{\it p_3}&{\it n_1}&0\\ \noalign{\medskip}-{
\it f_2}&-{\it k_1}&-{\it m_2}&0&{\it p_2}&{\it p_4}&{\it t_1}&0
\\\hline \noalign{\medskip}-{\it s_2}&-{\it s_4}&{\it p_1}&-{\it p_2}&0&{\it c_2}
&{\it q_1}&0\\ \noalign{\medskip}-{\it s_4}&{\it s_2}&{\it p_3}&-{\it p_4}&
-{\it c_2}&0&{\it q_2}&0\\ \hline\noalign{\medskip}{\it d_1}&{\it r_1}&-{\it n_1}
&-{\it t_1}&-{\it q_1}&-{\it q_2}&0&0\\ \noalign{\medskip}0&0&0&0&0&0&0&0
\end {array} \right).
   \\
 \end{eqnarray}
\section{conclution}
We have obtained the algebraic structures of  $N=(4,4)$ and  $N=(8,8)$ SUSY two dimensional sigma models on Lie groups (in general) and  $N=(4,4)$ and $N=(8,8)$ SUSY WZW models (in special). We have shown that as for  $N=(2,2)$ SUSY WZW case these structures correspond to the Lie bialgebra structures with one 2-cocycle for  $N=(4,4)$  SUSY WZW and two 2-cocycles for  $N=(8,8)$ SUSY WZW case. $N=(4,4)$ structure on Lie algebra  is a composition of two $N=(2,2)$  algebraic bi-Hermitian structures with extended condition and $N=(8,8)$ structure on Lie algebraic (double bihypercomplex structure) is a composition of two $N=(4,4)$ structure (algebraic bihypercomplex structure) with extended condition, then we will have full knowledge from SUSY sigma models and WZW models if we have full knowledge of $N=(2,2)$ and $N=(4,4)$ structures. As an open problem one can use the relations of  algebraic structures for $N=(4,4)$ and $N=(8,8)$ ((\ref{25})-(\ref{31})) to obtain and classify all these structures on low dimensional Lie algebra as for the $N=(2,2)$ case \cite{RS}.
\par \textbf{Acknowledgment:}
We would like to thanks from  F. Darabi and M.Sephid for carefully reading the manuscript and useful comments.


\end{document}